\documentclass{article}
\usepackage{spconf,amsmath,graphicx}
\usepackage{amsmath,amsfonts,amssymb,pifont}
\usepackage{adjustbox}
\usepackage{comment}
\usepackage{hyperref}
\usepackage{svg,comment}
\usepackage{algorithmic}
\usepackage{array}
\usepackage[caption=false,font=normalsize,labelfont=sf,textfont=sf]{subfig}
\usepackage{textcomp}
\usepackage{stfloats}
\usepackage{url}
\usepackage{verbatim}
\usepackage{cite}
\usepackage{booktabs}
\usepackage{color}
\usepackage[font=footnotesize]{caption}
\usepackage{multirow,multicol}
\usepackage{color}
\usepackage[font=footnotesize]{caption}
\newcommand{\tabitem}{~~\llap{\textbullet}~~}
\usepackage{xparse}

\usepackage{multirow,makecell}

\title{Prompting audios using Acoustic Properties for Emotion Representation}
\name{Hira Dhamyal$^1$, Benjamin Elizalde$^2$, Soham Deshmukh$^2$, Huaming Wang$^2$ , Bhiksha Raj$^{1,3}$, Rita Singh$^1$ \vspace{-8.5pt}
\thanks{This work was done when the first author was an intern at Microsoft}}
\address{
    $^1$ Carnegie Mellon University,
    $^2$Microsoft,
    $^3$Mohammed bin Zayed University of AI}
\begin{document}
\ninept
\maketitle
\begin{abstract}
Emotions lie on a continuum, but current models treat emotions as a finite valued discrete variable. This representation does not capture the diversity in the expression of emotion. To better represent emotions we propose the use of natural language descriptions (or prompts). In this work, we address the challenge of automatically generating these prompts and training a model to better learn emotion representations from audio and prompt pairs. We use acoustic properties that are correlated to emotion like pitch, intensity, speech rate, and articulation rate to automatically generate prompts i.e. ‘acoustic prompts’. We use a contrastive learning objective to map speech to their respective acoustic prompts. We evaluate our model on Emotion Audio Retrieval and Speech Emotion Recognition. Our results show that the acoustic prompts significantly improve the model’s performance in EAR, in various Precision@K metrics. In SER, we observe a 3.8\% relative accuracy improvement on the Ravdess dataset. 
\end{abstract}

\section{Introduction}
\label{sec:intro}
Emotions are usually described using discrete labels like 'angry', or 'happy' following psychological models like the Plutchik wheel of emotion \cite{plutchik1991emotions} or Ekman's model of emotion \cite{ekman1992there}. Although these frameworks are extremely popular and provide ease of modeling, they do not fully capture the diversity in emotion expression. This makes using such discrete representations sub-optimal for downstream tasks. 

Understanding the source of diversity in emotion expression is the key to formulating more accurate emotion representations. There are many sources of diversity in emotion, like the speaker, culture, and context, among other factors \cite{memon2019detecting, dhamyal2020phonetic}. Labeling two instances of emotion with the same label of say `anger', ignores the intricacies of the expression of anger. Therefore, we believe it is important to represent the fine-grained characteristics of emotion. 

These fine-grained characteristics of emotions can be better captured by the flexibility that natural language provides. In general, such descriptions can describe the low-level information in the audio like the acoustic properties or they can describe the high-level information like who is expressing the emotion and what the context is. Humans often use affective language to casually describe emotion in speech, for example, `An angry man shouting loudly'. In this example `loudness' has a direct acoustic correlate `intensity' which can be used to form a description e.g. `this is the sound of high-intensity anger'.

The choice of natural language description affects the high dimensional representation learned from the text, hence it is very important to choose the right description for the emotion. This leads to the question:

\noindent \emph{How do we describe an emotion using natural language and how can a model learn it?} 

In this work, we propose a method to describe the emotion in audio by using the low-level information in the audio. Previous research shows that there are numerous acoustic correlates of emotion \cite{frick1985communicating, scherer1972acoustic, pavlenko2005emotions}. These acoustic correlates include measurements like the average pitch, intensity, speech rate, and articulation rate. We extract these correlates from each utterance and use them to form the description in an automatic and scalable way. We call descriptions generated in this manner \emph{`acoustic prompts'}. 

Given these acoustic prompts, we train models that associate them with corresponding audio by fine-tuning the Contrastive Language-Audio Pretraining (CLAP) model \cite{elizalde2022clap, deshmukh2022audio}. CLAP uses contrastive learning to associate the audio and their descriptions and yields state-of-the-art performance in learning audio concepts with natural language descriptions. 
We then evaluate this fine-tuned model on downstream tasks. 

We evaluate on Emotion Audio Retrieval (EAR) and Speech Emotion Recognition (SER).
SER is a well-known task defined as given a speech utterance, determine the emotion present in the utterance \cite{dhamyal2022positional, dhamyal2020phonetic}. The task of EAR is not a commonly performed task. There are tangential works e.g. \cite{wang2023research, thao2023emomv} which examine retrieval of music audios, however, this task has not been explored for speech emotion. We believe that EAR is an important task to address since it can be useful in speech forensics, recommendation systems, search engines, social media, etc. Since emotions are also indicators of certain events, EAR methods can help in retrieving hate speech, and violence from audio. We show that the acoustic prompts improve the model's performance in EAR significantly; Precision@$K$  is consistently better for various values of $K$. We also find that in SER, the model performance improves. Specifically, recognition performance improves $3.8\%$ relative on Ravdess dataset. In a fine-tuning classification setup, we observe $3.7\%$ improvement on Ravdess.

In summary, the contributions of this paper are as follows:
\begin{enumerate}
    \item In this work, we propose a unified framework to train emotion representation model through audio-text contrastive learning. We explore ways to generate emotion prompts for speech, grounded in acoustic properties of pitch, intensity, speech rate, and articulation rate. 
    \item We introduce the task of text-based audio retrieval for emotion (not done before as far as we know) and show that our proposed prompts significantly improve performance on this task.
    \item We show improvements in two tasks; SER and EAR on a model trained on multiple emotion datasets.
\end{enumerate}


\section{Background}


\begin{figure*}
    \centering
    \includegraphics[width=0.6\linewidth,height=0.25\linewidth]{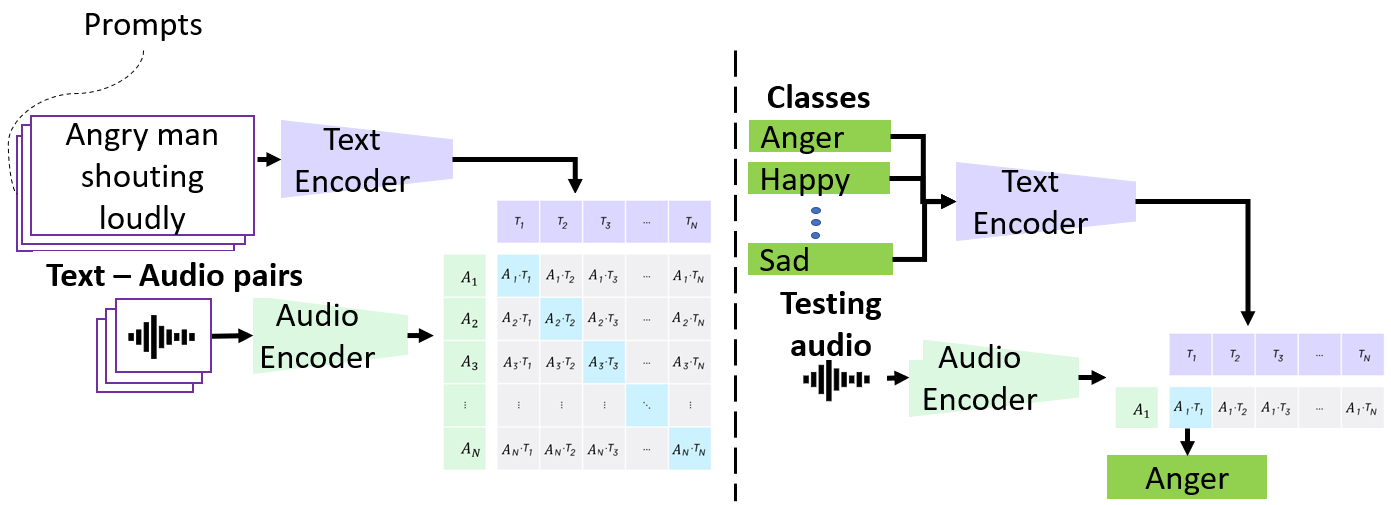}
    \caption{The left part of the image shows model training. Given a batch of $N$ audio-text pairs, the model trains the audio and text encoders to learn their (dis)similarity using contrastive learning. On the right side is shown an evaluation scenario. Given an audio of unknown emotion, trained audio and text encoders are used to extract representations from the audio and the descriptions. The prediction is made based on the cosine similarity between the two representations.  \vspace{-0.08in}}
    \label{fig:clap_model}
\end{figure*}

Fig.~\ref{fig:clap_model} shows the Contrastive Language-Audio Pretraining (CLAP) model - the backbone architecture used in this paper. The audio-text pairs are passed through an audio encoder and a text encoder respectively. Let $f_a(.)$ represent the audio encoder and $f_t(.)$ represent the text encoder. For a batch of N:
\begin{equation}
    \hat{X}_a = f_a(X_a); 
    \hat{X}_t = f_t(X_t)
\end{equation}
where $\hat{X}_a \in \mathbb{R}^{N \times V}$ are the audio representations of dimensionality $V$, and $\hat{X}_t \in \mathbb{R}^{N \times U}$ are the text representations of dimensionality $U$. 

We brought audio and text representations into a joint multimodal space of dimension $d$ by using a projection layer: 
\begin{equation}
    E_a = L_a(\hat{X}_a);
    E_t = L_t(\hat{X}_t)
\end{equation}
where $E_a \in \mathbb{R}^{N \times d}$, $E_t \in \mathbb{R}^{N \times d}$, $L_a$ and $L_t$ are the linear projections for audio and text respectively. 

Now that the audio and text embeddings ($E_a$, $E_t$) are comparable, we can measure similarity:
\begin{equation}
    C = \tau*(E_t \cdot E_a^\top)
\end{equation}
where $\tau$ is a temperature parameter to scale the range of logits. The similarity matrix $C \in \mathbb{R}^{N \times N}$ has $N$ correct pairs in the diagonal and $N^2-N$ incorrect pairs in the off-diagonal. The loss can be calculated as: 
\begin{equation}
     \mathcal{L} = 0.5 * (\ell_{text}(C) + \ell_{audio}(C))
\end{equation}
where $\ell_{k} = \frac{1}{N}\sum_{i=0}^{N} \log diag (softmax(C))$ along text and audio axis respectively. We used this symmetric cross-entropy loss ($\mathcal{L}$) over the similarity matrix to jointly train the audio and text encoders along with their linear projections. 

In this paper, we chose this architecture because it yields SoTA performance in learning audio concepts with natural language descriptions. We use log Mel spectrograms from the audios, sampled at 44K Hz, as input to the audio encoder - CNN14~\cite{kong2020panns}, which is pre-trained on 2M audio clips from AudioSet. The text encoder is BERT uncased. The audio encodings are of 1024 dimensional
from the HuggingFace library \cite{wolf2019huggingface}, 
whereas text encodings are 768 dimensional. Both encodings are then projected into a joint multimodal space of dimension 1024. 
Both audio and text encoders are frozen in our experiments, but the projection layers are learnable. 
We use PyTorch to implement the model architecture. The model is trained with 0.0001 learning rate, batch size of 128, for 30 epochs using Adam optimizer. 

\section{Proposed Work}
\subsection{Datasets}

We use 6 Emotion Datasets (ED) in this setup, see Table~\ref{tab:emotion_dataset}. The literature using these many datasets for emotion tasks are rare. The original CLAP model is trained with audio-text pairs sourced from three audio captioning datasets: ClothoV2~\cite{clotho}, AudioCaps~\cite{audiocaps}, MACS~\cite{macs}, and one sound event dataset: FSD50K~\cite{fsd50k}. Altogether they are referred to as 4D henceforth. All the datasets used are publicly available. 



\begin{table}
\footnotesize
\caption{Details of the 6 emotion datasets used in this paper.}
    \centering
    \begin{tabular}{lccl}
    \toprule
        Dataset&Files&Class&Emotions\\
        \midrule
        CMU-MOSEI\cite{zadeh2018multimodal}&\multirow{2}{0pt}{23K}&\multirow{2}{0pt}{9}&ang, exc, fear, sad\\ 
        & & &frus, neu, sur, hap, dis\\
        \hline
        IEMOCAP\cite{busso2008iemocap}&\multirow{2}{0pt}{10K}&\multirow{2}{0pt}{9}& hap, fear, sad, sur, exc, \\ 
        & & &ang, neu, disappoint, frus \\
        \hline
        MELD\cite{poria2018meld}&\multirow{2}{0pt}{10K}&\multirow{2}{0pt}{7}& neu, sur, fear, sad, \\ 
        & & &joy, disgust, ang \\
        \hline
        CREMA-D\cite{cao2014crema} & \multirow{2}{0pt}{7K}& \multirow{2}{0pt}{6} & ang, dis, fear, hap, \\
        & & &  neu, sad \\
        \hline
        RAVDESS\cite{livingstone2018ryerson} & \multirow{2}{0pt}{2.5K}& \multirow{2}{0pt}{8} & neu, calm, hap, sad, \\ 
        & & &  ang, fear, disgust, sur\\
        \hline
        CMU-MOSI\cite{zadeh2018multimodal} & $\qquad$ 2.2K  & $\mkern7mu$ 3 & neu, positive, negative \\
    \bottomrule
    \end{tabular}
    \label{tab:emotion_dataset}
\end{table}




\subsection{Prompt Generation}
\label{sec:prompt_gen}

For all the emotion datasets being used, we only have the discrete class labels no associated descriptions. Therefore, we devise a scalable and automatic prompting method that is based on the acoustic properties of the speech audios. There are numerous acoustic correlates of emotion therefore, we hypothesize that including this information in the prompts would benefit downstream emotion tasks. We construct the prompts in the manner described below: 



\noindent \textbf{Class label Prompt}
The simplest description for each audio can be the class label, i.e. audio with the discrete true label of `anger' will be labeled as `anger'. We use this as the baseline prompt to compare against the proposed prompts. 

\noindent \textbf{Pitch Prompt}
Pitch is known to be affected by emotion, lower pitch is related to negative emotions like fear and high pitch is related to positive emotions like happiness or surprise \cite{scherer1972acoustic}. We bin pitch into four bins, since pitch is naturally sex-specific i.e.  low-male pitch ($<132.5$ Hz), high-male pitch ($>132.5$ Hz,  $<180$ Hz), low-female pitch ($>180$ Hz, $<210$ Hz) and high-female pitch ($>210$ Hz)
However, we also experiment with binning into two classes, based on a cutoff of $170$ Hz. The cutoffs are obtained from the average numbers for vocal pitch reported in the literature \cite{traunmuller1995frequency}. The prompt is set as `{bin-class} {emotion-class}', an example of which is `low pitch anger' (without sex information) or `low male pitch anger' (otherwise). \\

\noindent \textbf{Intensity Prompt}
Intensity is known to be affected by emotion, low intensity is linked with negative emotions like sadness or melancholy and high intensity is linked with joy or excitement \cite{scherer1972acoustic}. We bin the average intensity over the audio clip in two bins, low and high intensity at $60$ dB \cite{roebuck2015attending}.
The cutoffs are based on average intensity numbers reported for human speech in literature. 
The same rule as pitch prompt is followed to form the intensity prompt, an example of which is `high intensity anger'.\\

\noindent \textbf{Speech-rate Prompt}
It has been observed that faster-spoken speech is linked with highly potent emotions such as anger and happiness whilst slower speech is linked with sadness, disgust, and boredom \cite{frick1985communicating}. Speech rate is calculated by extracting the number of syllables spoken divided by the total duration of the audio clip. We use $3.12$ syllables/sec as the cutoff to bin the speech rate into two bins, low and high speech rate \cite{martinez2016speech}. An example of a speech-rate prompt is `high speech rate anger'.\\

\noindent \textbf{Articulation-rate Prompt}
Similarly to speech rate, fast articulation rate is linked with emotions of interest, fear, or happiness; whereas slow articulation rate is indicative of sadness and disgust \cite{frick1985communicating}. The articulation rate is calculated as the total number of syllables divided by the total phonation time. We bin the audio into low and high articulation rate at the cutoff of $4$ syllables/sec \cite{martinez2016speech}.
An example of articulation-rate prompt is `high articulation rate anger'. \\ 
Even though speech and articulation rate are similar concepts, speech rate captures speaker-specific information in the form of the number of pauses and hesitation whereas articulation rate ignores such information.

\noindent \textbf{Prompt Augmentation}
To combine all 5 prompts, we pair an audio clip independently with each acoustic prompt. Thus, one audio clip will result in 5 pairs used for training our model.  
Note: we also tried making one prompt with all the acoustic properties combined together. However, this does not perform as well as when the prompts are paired separately with a given audio. 

\begin{table}[]
    \footnotesize
    \centering
    \caption{Given audio of class label \{emotion\}, the prompts generated will be one among the following.}
    \begin{tabular}{l|l}
    \toprule
    Property & Prompt \\
    \hline
        Class label (CL) & \tabitem \{emotion\} \\
        \hline 
        Pitch & \tabitem high female pitch \{emotion\}\\
        &       \tabitem low female pitch \{emotion\}\\ 
        &       \tabitem high male pitch \{emotion\}\\
        &       \tabitem low male pitch \{emotion\}\\
        \hline
        Intensity & \tabitem high intensity \{emotion\} \\
        &        \tabitem low intensity \{emotion\} \\
        \hline
        Speech rate & \tabitem high speech rate \{emotion\} \\
        &       \tabitem low speech rate \{emotion\} \\
        \hline
        Articulation rate & \tabitem high articulation rate \{emotion\} \\
        &   \tabitem low articulation rate \{emotion\} \\
        \bottomrule
    \end{tabular}
    \label{tab:prompt_examples}
\end{table}

Table \ref{tab:prompt_examples} shows all the acoustic prompts that are used in this work. 
We calculate the pitch and intensity using Librosa \cite{mcfee2015librosa} and we calculate speech rate and articulation rate using Praat \cite{praat}. 
Note: Other methods to select thresholds (used in prompt creation) like dataset-specific thresholds showed little effect on the final results, therefore we choose to use the literature-inspired thresholds.

\section{Experiments and Results}

\begin{table*}[]
\footnotesize
    \centering
    \caption{Precision@$K$ achieved under different training conditions and prompt settings. 
    The rows show three different models. The first row is the baseline CLAP model. The second and third rows are models trained on 5 emotion datasets, not including the IEMOCAP dataset. The second row is when the prompts used for training are the emotion class labels (CL) of the audios and the third row is when the prompts are acoustic prompts. PA refers to Prompt-Augmentation
    The queries here are the acoustic prompts also shown in Table \ref{tab:prompt_examples}. The model trained with acoustic prompt augmentation (PA) is consistently better.}
    \begin{tabular}{c|ccc|ccc|ccc|ccc|ccc}
    \toprule
         & \multicolumn{3}{c}{Class Label Queries} & \multicolumn{3}{c}{Pitch Queries} & \multicolumn{3}{c}{Intensity Queries} & \multicolumn{3}{c}{Speech Rate Queries} & \multicolumn{3}{c}{Articulation Rate Queries}\\
                       & P@1 & P@5 & P@10  & P@1 & P@5 & P@10 & P1 & P@5 & P@10 & P@1 & P@5 & P@10  & P@1 & P@5 & P@10 \\
    \midrule
        4D   &        0.50 & 0.35 & 0.35      &         0.07 & 0.12 & 0.10 & 0.13 & 0.18 & 0.19 & 0.25 & 0.20 & 0.15 & 0.13 & 0.10 & 0.14  \\
        4D + [5 ED - \textit{CL}] & 0.50 & 0.40 & 0.35 & 0.00&	0.04	&0.05&	0.25&	0.13&	0.15	&0.13&	0.18&	0.19&	0.13&	0.13&	0.13 \\

        4D + [5 ED - \textit{PA}] & \textbf{0.75} & \textbf{0.45} & \textbf{0.38} & \textbf{0.20} &	\textbf{0.13} &	\textbf{0.15}&	\textbf{0.25}&	\textbf{0.20}&	\textbf{0.20}&\textbf{0.38}&	\textbf{0.25}	&\textbf{0.23}&	\textbf{0.38}&	\textbf{0.23}&	\textbf{0.21}\\
    \bottomrule
    \end{tabular}
    \label{tab:acoustic_prompt_retrieval}
\end{table*}

\subsection{Emotion Audio Retrieval} \label{ear_section}
We evaluate our trained models for the task of emotion audio retrieval (EAR). With the increasing sizes of audio databases, being able to search such databases for specific types of audio is important. We compare (1) the baseline CLAP model, (2) the model where the prompts used for training are the emotion class labels of the audio, and (3) the model trained with our acoustic prompting method using prompt augmentation.

The first three columns in Table \ref{tab:acoustic_prompt_retrieval} show the results when the queries are among the four emotion classes, i.e. happy, sad, angry, and neutral and the collection consists of IEMOCAP dataset. Row 1 model is trained on only 4 audio captioning datasets. Rows 2 and 3 models are trained on 5 emotion datasets, not including IEMOCAP. For a given query, the model outputs top $K$ audios whose audio embeddings have the highest cosine similarity to the text embedding of the query. 

We observe that the model trained on acoustic prompts performs significantly better for all the precision@$K$ metrics. This shows that training the model with acoustic prompts is resulting in better-learned emotion representations. 

Furthermore, we also access whether the trained model learns associations between the acoustic properties and the speech emotion. We test this in a similar framework as in the last experiment. 
The queries are made similar to the prompts as shown in Table \ref{tab:prompt_examples}. 

The rest of the columns in Table \ref{tab:acoustic_prompt_retrieval} show the results of audio retrieval when queries are from the acoustic prompts. We calculate precision@$K$ for each acoustic prompt shown on the columns. 
From the results, we observe that the model trained on the proposed acoustic prompting method performs best in all cases. 
The takeaway here is that our model is able to retrieve audio significantly better when trained using acoustic prompt augmentation. The precision@$K$ numbers are comparable to numbers observed in audio retrieval tasks \cite{kim2019improving}. 
The results suggest that we can introduce even more elaborate descriptions for each audio at training time and the model will learn associations and be able to retrieve audios with those descriptions.


\subsection{Speech Emotion Recognition} \label{ser_section}
\begin{table}[]
\caption{Accuracy \% on Ravdess when the model is trained under different settings. The second column shows when Ravdess is not in the training sets. The third column shows when the model is finetuned on Ravdess. The second row shows the CLAP Baseline trained on 4 audio captioning datasets (4D). Third row is when the model is trained using only 5 Emotion Datasets (5 ED). The following rows include 4D and 5ED in training and for the ED, the prompts during training are either the class labels (CL) or the acoustic prompt augmentation (PA) respectively.}
    \centering
    \begin{tabular}{lc|c}
         \toprule
         Training dataset  & Leave one out & Finetune \\
         \midrule
         Random & 12.50 &  12.50 \\
         4D &	15.99 &	68.50 \\
        5 ED - \textit{CL} & 	22.88 & 68.50 \\
        4D + [5ED - \textit{CL}]	&	\textbf{38.46}  &	68.69\\
        4D + [5ED - \textit{PA}]	&	27.88 & \textbf{72.46}\\
        SoTA & - & 81.82 \cite{luna2021proposal} \\
        \bottomrule
        \vspace{-1cm}
    \end{tabular}
    \label{tab:ood_supervised_results}
\end{table}
To evaluate how the acoustic prompts would help in SER, we perform the following two experiments. The first is a zero-shot like setup where we leave one dataset out, which is used during the testing stage. The second is a fine-tuning setup where the model from the first setup is fine-tuned on the left-out dataset. 

\subsubsection{Leave one out}
This setup evaluates how well a model trained on a pre-defined set of classes generalizes to a new dataset, which might have same or different sets of classes. 
Out of the 6 emotion datasets, we leave one out for testing and train the model on the other 5 emotion datasets. Therefore the training and testing datasets are completely different. In the case where Ravdess is the testing dataset, `calm' class is not represented in any of the other training datasets and is a zero-shot classification result. 

We train 5 different models shown in the rows of Table~\ref{tab:ood_supervised_results}. There are two main takeaways from this experiment. Firstly adding Emotion datasets in the training stage helps the performance on the left-out emotion dataset. This can be observed in the second column where the performance improves from $15.99\%$ to $22.88\%$. 

Secondly using acoustic prompt augmentation (PA) is not helping in the fine-tuning setup. We believe this is because there is a distribution shift in the training and testing datasets, which effects the acoustics and hence the acoustic prompts. For example, `high intensity anger' prompt might not be prevalent in the training datasets but is present in the testing dataset. This harms the transferability of the learned acoustic prompts to a completely new dataset. 
Note that the SoTA performance for this evaluation setup is not found in literature because the general evaluation setup is when the dataset is present in both training and testing sets.


\subsubsection{Finetune}
In this experiment, we fine-tune the model from the previous stage on the left-out dataset. 

The results for SER are shown in the last column of Table~\ref{tab:ood_supervised_results}. 
We observe that when using acoustic prompt augmentation, we get the best accuracy metric. We see improvement in performance by absolute $3.77\%$, from $68.69\%$ to $72.46\%$.

\subsection{Prompt Analysis}
To evaluate which of the proposed acoustic prompts is better, we apply the trained model on SER with a smaller setup as in the last experiment, where the testing dataset is present in the training dataset. 
The model is trained 6 different times, where each time the description associated with emotion audios are varied. Among the 6, 1 uses the class label prompt and 4 uses the acoustic prompts as described in Section \ref{sec:prompt_gen}, and 1 uses the prompt augmentation - which combines all the acoustic prompts. 

\begin{figure}[tbh!]
  \centering
  \vspace{-0.3cm}
  \includegraphics[width=0.45\linewidth,height=0.25\linewidth]{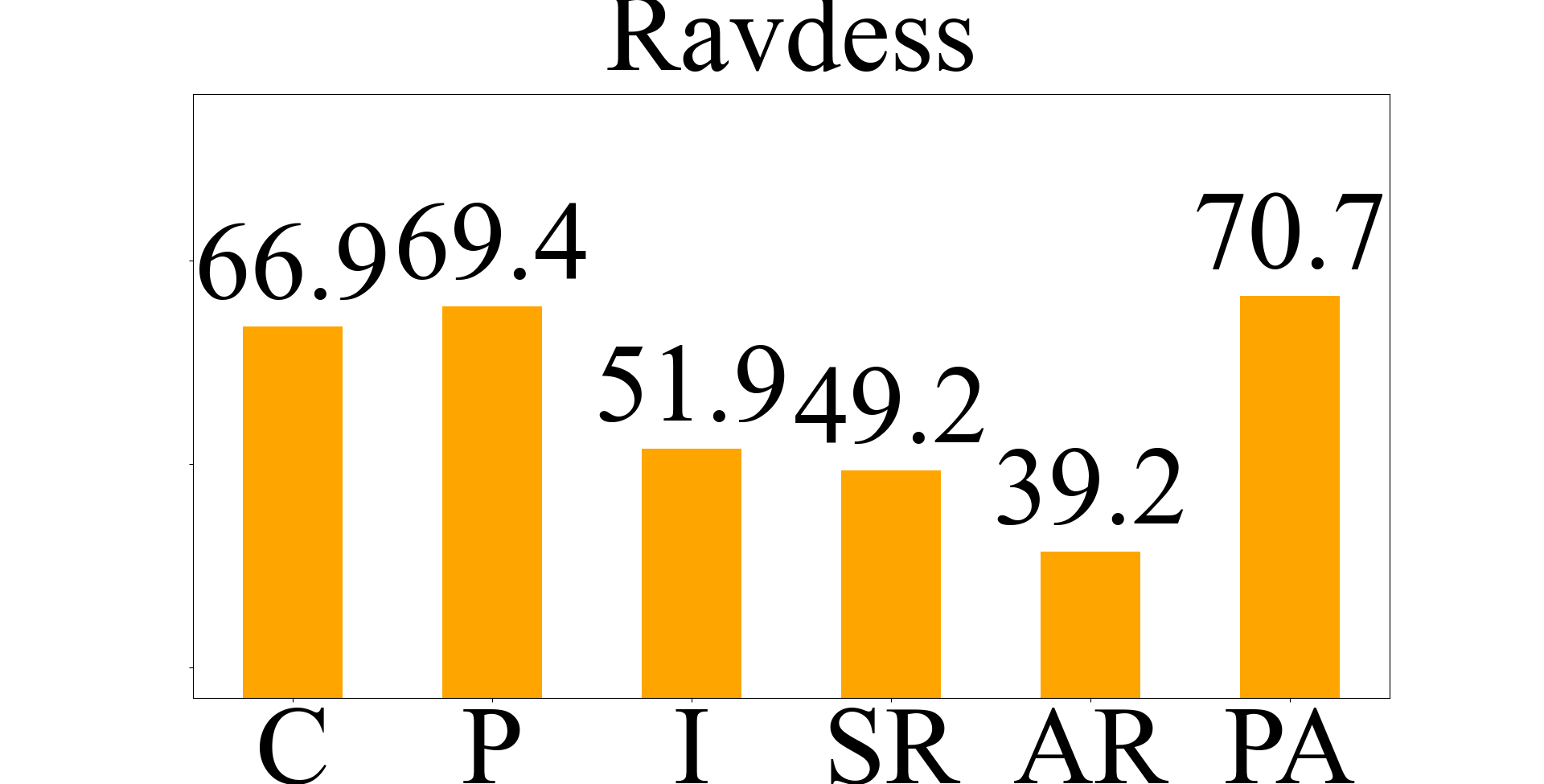}
  \caption{Accuracy achieved using different acoustic prompts on Ravdess. C=Class label, P=Pitch prompt, I=Intensity prompt, SR=Speech-Rate prompt, AR=Articulation-Rate prompt, PA=Prompt Augmentation.}
    \label{fig:prompt_analysis}
          \vspace{-0.1in}
\end{figure}
We train the model on 4 audio captioning datasets and 1 emotion dataset. The left part of Figure \ref{fig:prompt_analysis} shows the performance achieved when the model is trained on the training set (including 4D and Ravdess) and tested on the testing set of Ravdess.
We observe that among the 4 acoustic prompts, the pitch prompt gives the best performance. 
The second-best performance is achieved by the intensity prompt, followed by speech rate and then articulation rate.
Secondly, we observe that overall acoustic prompt augmentation is giving the best performance in both datasets. 

\section{Limitations and Conclusion}
There are certain limitations to our work.  Firstly, we use only four acoustic properties, however, there are other acoustic properties that are effected by emotion and should be explored. Secondly for each prompt, we create 2 or 4 bins per acoustic property, while these bins could be more fine-grained. Our future study will include work in alleviating the need for thresholding and relying on data-centric methods of binning the prompts. 

This work performs SER and EAR using the audios and their automatically generated descriptions. We use the acoustics of emotions to prompt the audios, in fact, there can be more complicated descriptions, invoking the semantics, environment, and context among other factors. We envision that as methods of describing emotions become more complicated, our ability to model emotions will become better. 
The acoustic properties we extract include pitch, intensity, speech rate, and articulation rate extracted from the audio.
We find that among the acoustic prompts, pitch prompt is the best performing. Overall for EAR when we do acoustic prompt augmentation, we achieve consistently better Precision@K metric. For SER, we also achieve an improvement in performance in Ravdess by $3.8\%$ in the finetuning setup. 

\pagebreak

\bibliographystyle{IEEEbib}
\bibliography{refs}

\begin{thebibliography}{10}

\bibitem{plutchik1991emotions}
Robert Plutchik,
\newblock {\em The emotions},
\newblock University Press of America, 1991.

\bibitem{ekman1992there}
Paul Ekman,
\newblock {\em Are there basic emotions?},
\newblock American Psychological Association, 1992.

\bibitem{memon2019detecting}
Shahan~Ali Memon, Hira Dhamyal, Oren Wright, Daniel Justice, Vijaykumar Palat, William Boler, Bhiksha Raj, and Rita Singh,
\newblock ``Detecting gender differences in perception of emotion in crowdsourced data,''
\newblock {\em arXiv preprint arXiv:1910.11386}, 2019.

\bibitem{dhamyal2020phonetic}
Hira Dhamyal, Shahan~Ali Memon, Bhiksha Raj, and Rita Singh,
\newblock ``The phonetic bases of vocal expressed emotion: Natural versus acted,''
\newblock {\em Proc. Interspeech 2020}, pp. 3451--3455, 2020.

\bibitem{frick1985communicating}
Robert~W Frick,
\newblock ``Communicating emotion: The role of prosodic features.,''
\newblock {\em Psychological bulletin}, vol. 97, no. 3, pp. 412, 1985.

\bibitem{scherer1972acoustic}
Klaus~R Scherer,
\newblock ``Acoustic concomitants of emotional dimensions: Judging affect from synthesized tone sequences.,''
\newblock 1972.

\bibitem{pavlenko2005emotions}
Aneta Pavlenko,
\newblock {\em Emotions and multilingualism.},
\newblock Cambridge University Press, 2005.

\bibitem{elizalde2022clap}
Benjamin Elizalde, Soham Deshmukh, Mahmoud~Al Ismail, and Huaming Wang,
\newblock ``Clap: Learning audio concepts from natural language supervision,''
\newblock {\em arXiv preprint arXiv:2206.04769}, 2022.

\bibitem{deshmukh2022audio}
Soham Deshmukh, Benjamin Elizalde, and Huaming Wang,
\newblock ``Audio retrieval with wavtext5k and clap training,''
\newblock {\em arXiv preprint arXiv:2209.14275}, 2022.

\bibitem{dhamyal2022positional}
Hira Dhamyal, Bhiksha Raj, and Rita Singh,
\newblock ``Positional encoding for capturing modality specific cadence for emotion detection,''
\newblock {\em Proc. Interspeech 2022}, pp. 166--170, 2022.

\bibitem{wang2023research}
Weixing Wang, Qianqian Li, Jingwen Xie, Ningfeng Hu, Ziao Wang, and Ning Zhang,
\newblock ``Research on emotional semantic retrieval of attention mechanism oriented to audio-visual synesthesia,''
\newblock {\em Neurocomputing}, vol. 519, pp. 194--204, 2023.

\bibitem{thao2023emomv}
Ha~Thi~Phuong Thao, Gemma Roig, and Dorien Herremans,
\newblock ``Emomv: Affective music-video correspondence learning datasets for classification and retrieval,''
\newblock {\em Information Fusion}, vol. 91, pp. 64--79, 2023.

\bibitem{kong2020panns}
Qiuqiang Kong, Yin Cao, Turab Iqbal, Yuxuan Wang, Wenwu Wang, and Mark~D Plumbley,
\newblock ``Panns: Large-scale pretrained audio neural networks for audio pattern recognition,''
\newblock {\em IEEE/ACM Transactions on Audio, Speech, and Language Processing}, vol. 28, pp. 2880--2894, 2020.

\bibitem{wolf2019huggingface}
Thomas Wolf, Lysandre Debut, Victor Sanh, Julien Chaumond, Clement Delangue, Anthony Moi, Pierric Cistac, Tim Rault, R{\'e}mi Louf, Morgan Funtowicz, et~al.,
\newblock ``Huggingface's transformers: State-of-the-art natural language processing,''
\newblock {\em arXiv preprint arXiv:1910.03771}, 2019.

\bibitem{clotho}
Konstantinos Drossos, Samuel Lipping, and Tuomas Virtanen,
\newblock ``Clotho: an audio captioning dataset,''
\newblock in {\em IEEE International Conference on Acoustics, Speech and Signal Processing (ICASSP)}, 2020.

\bibitem{audiocaps}
Chris~Dongjoo Kim, Byeongchang Kim, Hyunmin Lee, and Gunhee Kim,
\newblock ``{AudioCaps: Generating Captions for Audios in The Wild},''
\newblock in {\em NAACL-HLT}, 2019.

\bibitem{macs}
Irene Mart{\'\i}n-Morat{\'o} and Annamaria Mesaros,
\newblock ``What is the ground truth? reliability of multi-annotator data for audio tagging,''
\newblock in {\em 2021 29th European Signal Processing Conference (EUSIPCO)}. IEEE, 2021, pp. 76--80.

\bibitem{fsd50k}
Eduardo Fonseca, Xavier Favory, Jordi Pons, Frederic Font, and Xavier Serra,
\newblock ``Fsd50k: An open dataset of human-labeled sound events,''
\newblock {\em IEEE/ACM Transactions on Audio, Speech, and Language Processing}, 2022.

\bibitem{zadeh2018multimodal}
AmirAli~Bagher Zadeh, Paul~Pu Liang, Soujanya Poria, Erik Cambria, and Louis-Philippe Morency,
\newblock ``Multimodal language analysis in the wild: Cmu-mosei dataset and interpretable dynamic fusion graph,''
\newblock in {\em Proceedings of the 56th Annual Meeting of the Association for Computational Linguistics (Volume 1: Long Papers)}, 2018, pp. 2236--2246.

\bibitem{busso2008iemocap}
Carlos Busso, Murtaza Bulut, Chi-Chun Lee, Abe Kazemzadeh, Emily Mower, Samuel Kim, Jeannette~N Chang, Sungbok Lee, and Shrikanth~S Narayanan,
\newblock ``Iemocap: Interactive emotional dyadic motion capture database,''
\newblock {\em Language resources and evaluation}, vol. 42, no. 4, pp. 335--359, 2008.

\bibitem{poria2018meld}
Soujanya Poria, Devamanyu Hazarika, Navonil Majumder, Gautam Naik, Erik Cambria, and Rada Mihalcea,
\newblock ``Meld: A multimodal multi-party dataset for emotion recognition in conversations,''
\newblock {\em arXiv preprint arXiv:1810.02508}, 2018.

\bibitem{cao2014crema}
Houwei Cao, David~G Cooper, Michael~K Keutmann, Ruben~C Gur, Ani Nenkova, and Ragini Verma,
\newblock ``Crema-d: Crowd-sourced emotional multimodal actors dataset,''
\newblock {\em IEEE transactions on affective computing}, vol. 5, no. 4, pp. 377--390, 2014.

\bibitem{livingstone2018ryerson}
Steven~R Livingstone and Frank~A Russo,
\newblock ``The ryerson audio-visual database of emotional speech and song (ravdess): A dynamic, multimodal set of facial and vocal expressions in north american english,''
\newblock {\em PloS one}, vol. 13, no. 5, pp. e0196391, 2018.

\bibitem{traunmuller1995frequency}
Hartmut Traunm{\"u}ller and Anders Eriksson,
\newblock ``The frequency range of the voice fundamental in the speech of male and female adults,''
\newblock {\em Unpublished manuscript}, vol. 11, 1995.

\bibitem{roebuck2015attending}
Hettie Roebuck, Kun Guo, and Patrick Bourke,
\newblock ``Attending at a low intensity increases impulsivity in an auditory sustained attention to response task,''
\newblock {\em Perception}, vol. 44, no. 12, pp. 1371--1382, 2015.

\bibitem{martinez2016speech}
F~Mart{\'\i}nez-S{\'a}nchez, JJG Meil{\'a}n, J~Carro, C~G{\'o}mez {\'I}{\~n}iguez, L~Millian-Morell, IM~Pujante Valverde, T~L{\'o}pez-Alburquerque, and DE~L{\'o}pez,
\newblock ``Speech rate in parkinson's disease: A controlled study,''
\newblock {\em Neurolog{\'\i}a (English Edition)}, vol. 31, no. 7, pp. 466--472, 2016.

\bibitem{mcfee2015librosa}
Brian McFee, Colin Raffel, Dawen Liang, Daniel~P Ellis, Matt McVicar, Eric Battenberg, and Oriol Nieto,
\newblock ``librosa: Audio and music signal analysis in python,''
\newblock in {\em Proceedings of the 14th python in science conference}, 2015, vol.~8, pp. 18--25.

\bibitem{praat}
``Praat,'' \url{https://www.fon.hum.uva.nl/praat}.

\bibitem{kim2019improving}
Bongjun Kim and Bryan Pardo,
\newblock ``Improving content-based audio retrieval by vocal imitation feedback,''
\newblock in {\em ICASSP 2019-2019 IEEE International Conference on Acoustics, Speech and Signal Processing (ICASSP)}. IEEE, 2019, pp. 4100--4104.

\bibitem{luna2021proposal}
Cristina Luna-Jim{\'e}nez, Ricardo Kleinlein, David Griol, Zoraida Callejas, Juan~M Montero, and Fernando Fern{\'a}ndez-Mart{\'\i}nez,
\newblock ``A proposal for multimodal emotion recognition using aural transformers and action units on ravdess dataset,''
\newblock {\em Applied Sciences}, vol. 12, no. 1, pp. 327, 2021.

\end{thebibliography}

\clearpage

\end{document}